# Crucial role of fragmented and isolated defects in persistent relaxation of deeply supercooled water

**Shinji Saito[1†], Biman Bagchi[2‡], and Iwao Ohmine[1§]**

[1] Institute for Molecular Science, Myodaiji, Okazaki, Aichi, 444-8585, Japan, and
The Graduate University for Advanced Studies, Myodaiji, Okazaki, Aichi, 444-8585, Japan

[2] Indian Institute of Science, Bangalore, 560012, India

Properties of water have been well elucidated for temperatures above ~230 K and yet mysteries remain in the deeply supercooled region. By performing extensive molecular dynamics simulations on this supercooled region, we find that structural and dynamical instabilities are hidden in the experimentally inaccessible region between 235 K and 150 K. We find a hitherto undiscovered fragmentation from 220 K to 190 K, which is the break-up of large clusters consisting of molecules with locally distorted tetrahedral structure into small pieces with one or two isolated defects. The fragmentation leads to considerable changes in the relaxation dynamics of water. We reveal a crucial role of specific three-coordinated defects in slow but persistent structural relaxation. The presence of relaxation due to these specific defects makes water glass transition temperature $T_g$ (= 136 K) extremely low and explains why the $T_g$ of water is ~1/2 of the melting temperature $T_m$, much lower than the commonly obeyed 2/3 rule of $T_g/T_m$.

Corresponding authors:
†E-mail: shinji@ims.ac.jp
‡E-mail: profbiman@gmail.com
§E-mail: ohmine@ims.ac.jp

# I. INTRODUCTION

Water is the most abundant liquid on earth and has numerous anomalous physio-chemical properties that play crucial roles in nature.[1, 2] Intensive studies have been made to elucidate anomalous properties of water by exploring supercooled state where their origins are considered to be imbedded.[3-15] Water has been investigated experimentally down to 235 K at ambient pressure. Below this temperature pure bulk water undergoes spontaneous crystallization and becomes hexagonal ice Ih. As for water properties in the supercooled state below 235 K, experimental studies have been carried out on water in droplets,[16, 17] in a confined system,[18] and in a hydrated protein system,[19] where water at low temperatures can be partially accessed by avoiding the formation of ice. Despite of these continuous attempts, the overall picture of supercooled water in this region remains to be clarified further.

To complement this experimental difficulty, computational studies have been performed by employing various water model potentials. The liquid-liquid critical point (LLCP) scenario was proposed[6] and confirmed with many water model potentials:[20-27] large local density fluctuations between the low-density liquid (LDL) with locally well-formed tetrahedral hydrogen bond (HB) structure and the high-density liquid (HDL) with locally distorted tetrahedral structure are found near the Widom line located between 220 K and 230 K at 1 atm[22-24, 27-30] (228 K obtained by extrapolating experimental data[9]). Furthermore, the crossover of population of HDL and LDL causes divergence-like behavior of thermodynamic response functions. Note that the presence of HDL and LDL is originated from the polyamorphism of water that exists in the form of low-density amorphous (LDA),[31, 32] high-density amorphous (HDA),[31, 32] and very high-density amorphous (VHDA) phases.[33] The detailed properties of supercooled water far below the temperature of Widom line have remained unclear even in theoretical and computational studies due to the computational difficulty of slow relaxation.

The glassy state of water has unique properties,[34] for example, the glass transition temperature, $T_g$, of water is distinctively low. For the most of glass-former materials, the glass transition temperature is found to be approximately 2/3 of the melting temperature, $T_m$, which has been known as the "2/3 rule" of $T_g/T_m$.[35] For water at ambient pressure, the glass transition temperature between LDL and LDA is 136K that is almost the half of the melting temperature of ice. The origin of this very low $T_g$ of water is not known yet. Upon further heating above $T_g$, a highly viscous LDL undergoes a crystallization to cubic ice Ic at 150 K. In the temperature range between 235 K and 150 K water cannot exist as a supercooled liquid state, i.e., it is always in an ice phase,[34, 36, 37] and thus this temperature region has been called "no man's land".[3] Recently the deeply supercooled water between 136 K and 150 K has been classified as an "extremely strong" liquid.[34] But the precise mechanism of its mobility and dynamics remains unknown and needs to be explained.

In the present study, we examine the structure and dynamics of liquid water from ambient to deeply supercooled conditions by performing extensive molecular dynamics (MD) simulations. We find that there exists a structural and dynamical transition at ~190 K where the fragmentation of locally distorted-structures takes place and the temperature dependence of relaxation time changes. Water changes its character from a fragile to a strong liquid. This transition at ~190 K is different from the well-known one associated with the crossing of the Widom line at ~220 K (228 K in experiment). We also find that specific three-coordinated defect molecules with one HB donor and two HB acceptors start to play a crucial role in the structural relaxation of deeply supercooled water.

## II. COMPUTATIONAL DETAILS

We carried out very long MD simulations under both NVE and NPT conditions with our own program and GROMACS.[38] The system size was 1000 water molecules. The periodic boundary condition was employed and the long-range electric interactions were calculated by using the Ewald sum. The equations of motion were solved with a time step of 2 fs. The TIP4P/2005 model potential, which is known to reproduce various water properties well, was used for water molecules.[39]

First, MD simulations under NPT conditions were performed to determine the densities at 22 temperatures: 300, 280, 275, 270, 250, 240, 235, 230, 225, 222, 220, 215, 205, 202, 197, 190, 180, 170, 160, 150, 145, and 130 K. The Nóse-Hoover thermostat and Berendsen barostat were used in the NPT simulations. Lengths of the NPT MD simulations were 50 ns, > 250 ns, and 500 ns for above 230 K, from 220 K to 170 K, and below 160 K, respectively. The calculated densities plotted in Fig. 1a are in good agreement with the result by Pi et al.[40]

Various structural and dynamical properties were analyzed by performing multiple NVE MD simulations at the determined average density above 130 K. The details are as follows. Three, five, and six independent trajectories were run above 180 K, at 180 K, and below 180K, respectively. For these simulations we first carried out the equilibration runs at 300 K at the density determined with the NPT simulations, which were then followed by the production runs. The last data in the equilibration runs at 300 K were used for the simulations at 280, 275, and 270 K. Below 270K, the last data in the equilibration runs at a high temperature were used to generate a new simulation at a next lower temperature; for example, the last data of the equilibration runs at 270 K were employed for the MD simulations at 250 K. This procedure was repeated for simulations down to 145 K.

For 130 K, one long MD simulation with 1.5 $\mu$s was performed using the data at 145 K. The lengths of equilibration and production runs at each temperature are presented in Table S1 in supplementary material; three sets of production runs at each temperature for 300 K – 190 K (the totally 45 ns production run at 300 K to that of 30 μs at 190 K), and five sets of 80 μs-production runs at 180 K and six sets at each temperature lower below 180 K (totally 400 μs production run at 180 and 480 μs at each temperature for 170 K – 145 K).

## III. RESULTS AND DISCUSSIONS

### A. Thermodynamic properties

The temperature dependences of specific heat $C_P$, compressibility $\kappa_T$, density $\rho$, and the temperature derivative of density $d\rho/dT$ are plotted in **Fig. 1**. As known from previous studies,[24, 27, 29, 30, 41, 42] Both of the $C_P$ and $\kappa_T$ exhibit maxima due to the crossover of HDL and LDL at 220-230 K corresponding to the locus of Widom line at the ambient pressure. With decreasing temperature from 220 K, the thermodynamic response functions first decrease sharply down to 190K and then gradually below 190 K. We note that the density has a maximum at ~270 K (277 K in experiment) and a minimum at ~190 K. *The temperature dependence of these properties seems to suggest that certain structural and dynamical changes take place at ~190, 220, and 270 K.*

It is noted that no crystallization has been observed in any of the present simulations; the total number of double-diamond cage units and hexagonal cage units, which form the basic building blocks for cubic and hexagonal ices,[43-45] never exceeds five, i.e., ~5% of total water molecules in the system, below 190 K.

We make a rough estimate that the crystallization time is more than $10^4$ times slower than the relaxation time at 200 K (see supplementary material). A recent study with the TIP4P/Ice model[46] also estimates that the minimum of crystallization time is approximately 10 μs at around 215 K (mentioned as 55 K below the melting point), whereas the relaxation time for molecular motions is a few tens ns.[47]

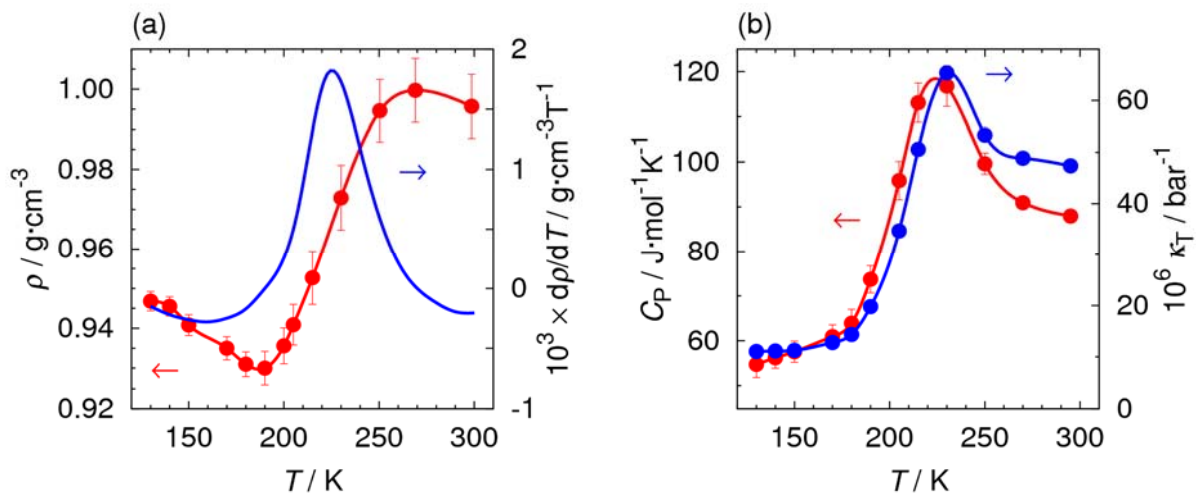

FIG. 1. Temperature dependencies of (a) the density (red) and its temperature-derivative (blue) and (b) the isobaric specific heat (red) and the isothermal compressibility (blue).

## B. Fragmentation and isolation of clusters in water

We examine the local liquid structure in water in terms of two kinds of local HB structures detailed below, based on our previous study.[29] Molecules without any HB defects within their second hydration shells, i.e., all the molecules up to their second hydration shells are four-coordinated, are defined as locally tetrahedral-structured molecules. And all other molecules are defined as locally distorted-structured molecules (explained in **Fig. S2a**). Although any local density is not used for the definition of two local structures, the local densities of locally tetrahedral- and distorted-structured molecules calculated from the volume of Voronoi polyhedra are 0.947 and 0.971 g/cm$^3$ at 220 K (0.942 and 0.963 g/cm$^3$ at 180 K), respectively. Therefore, the locally tetrahedral- and distorted-structured molecules can be regarded as the representative molecules in LDL and HDL, respectively. We note that although different classifications of two states have been employed to describe water properties in other studies, they essentially serve the same purpose of distinguishing between the two states.[28, 48-52] Our definition here allows a microscopic specification that proves useful, as shown later.

**Figure 2a** shows the temperature dependence of the fractions of locally tetrahedral-structured molecules and locally distorted-structured molecules. *The fraction of locally distorted-structured molecules is more than 90% at 300 K, and gradually decreases with lowering temperature (Fig. 2d).* According to our definition, most of the four-coordinated molecules are classified as the *locally* distorted-structured molecules since they have HB defects in their neighbors though the fraction of four-coordinated molecules itself is approximately 55% at 300 K (see **Fig. 4a**). Therefore, our classification allows a sharper delineation of the changing structures as we lower the temperature.

Between ~250 K and ~200 K, the large clusters of locally tetrahedral- and distorted-structured molecules both coexist without large-scale phase separation as interweaved in spatial dimension (as would be in a bi-continuous phase). The interchange of water molecules between these clusters gives rise to pronounced local density fluctuations (**Fig. 2e**). The crossover between the fractions of locally tetrahedral- and distorted-structured molecules occurs at ~225 K, which is approximately the same temperature of maxima of $C_P$ and $\kappa_T$ (Fig. 1b). Below 225 K, the decrease in the fraction of locally distorted-structured molecules is rapid. This continues till 190 K where the density exhibits the minimum. Below 190 K, the number of the locally distorted-structured molecules decreases slowly, although the number of such molecules is already very few (**Fig. 2f**). Thus, the situations below 190 K and above 270 K are completely opposite, regarding the populations of the locally tetrahedral- and locally distorted-structured molecules. Here we use the geometric criteria given by Luzar and Chandler[53] for the definition of HB. It is noted that the fractions and numbers of the locally tetrahedral- and distorted-structured molecules and other results presented here do not show any significant change regardless of the definition of HB (see **Figs. S2b** and **S2c** in which the results using the different definition of HB by Raiteri et al.[54] are presented.).

**Figure 2b** shows that the numbers of the clusters of the locally distorted- and locally tetrahedral-structured molecules change drastically with temperature. Here, we define the cluster by HBs connection. Above 250 K, the system is percolated by a single large cluster of the locally distorted-structured molecules (**Fig. S2d**).[55, 56] The analysis of the size distribution and of the second moment of the clusters (Figs. S2d and S2h) [57] shows that the percolation threshold of the locally distorted-structured molecules is located at ~240 K. Below 220 K, however, the number of clusters of the locally distorted-structured molecules sharply increases and their *size* decreases, i.e., the percolated cluster of the locally distorted-structured molecules is fragmented when the population of these defects falls below the critical percolation probability.

The number of the locally distorted-structured clusters peaks around 190-200 K (Fig. 2b) where the fragmentation is almost complete, and then slowly decreases with the further lowering of temperature. The percolation threshold of the locally tetrahedral-structured molecules is estimated to be at ~205 K (Figs. S2e and S2h). Along with the fragmentation and shrinking of the locally distorted-structured clusters below 190 K, the number of three- and five-coordinated molecules, i.e., the defect molecules, in an individual locally distorted-structured cluster drops and converges to a value ~1.5 (**Fig. 2c**), i.e*., most of the distorted-structured clusters consist of only one or two defects.* These small distorted-structured clusters are scattered and imbedded in the large-scale tetrahedral-structured network. The density minimum at ~190 K (Fig. 1a) is attributed to the fact that the scattered locally distorted-structured molecules do not influence the volume change below 190 K. The normal contraction force thus takes over below 190 K, leading to the increase in density with decreasing temperature.

Another fragmentation occurs at higher temperature and is completed at ~270 K (see Fig. 2b), which is the temperature of the density maximum in the simulation. This fragmentation of the locally tetrahedral-structured cluster can be regarded as the counterpart of the fragmentation of the locally distorted-structured clusters observed at 190 K, and discussed above.

The fragmentations caused by the cluster instability bear a resemblance to the process of a spinodal decomposition.[58] These instabilities occur in the "supercritical" region of the second critical point. Thus, the present two fragmentations may be called "extended" spinodal decompositions associated with the Widom line, distinguished from a conventional spinodal decomposition. The fragmentation of locally distorted-structured clusters completes at the temperature of density minimum. A similar picture may be applied for the other fragmentation at 270 K where the density is maximum. It will be thus important to examine the pressure dependence of the loci of the extended spinodal lines for understanding the entire picture of water.[51, 52]

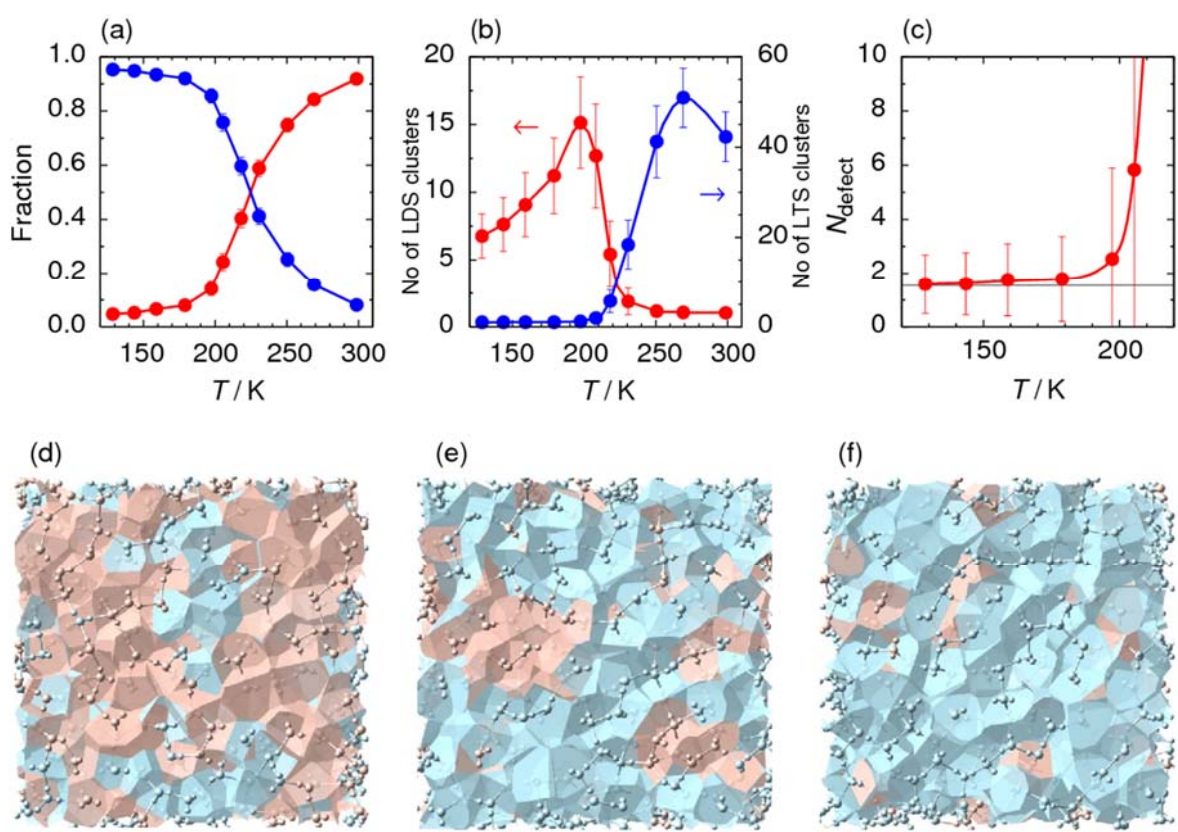

FIG. 2. Temperature dependences of (a) the fractions of locally tetrahedral-structured (blue) and locally distorted-structured (red) molecules, (b) the numbers of clusters of locally tetrahedral-structured (blue) and locally distorted-structured (red) molecules, and (c) the average number of HB defects per locally distorted-structured cluster. Snapshots of instantaneous structures at (d) 250 K, (e) 220 K, and (f) 180 K. The locally tetrahedral-structured and locally distorted-structured molecules are shown in blue and red in (d)-(f).

## C. Dynamical transitions and strong-fragile transition

We examine the temperature dependence of structural changes by calculating the self-term of intermediate scattering function (SISF),

$$F_s(k,t) = \frac{1}{N}\left\langle \sum_{j=1}^{N} \exp\left(i\mathbf{k}\cdot(\mathbf{r}_j(t)-\mathbf{r}_j(0))\right)\right\rangle \quad (1)$$

where $k$ is the wave vector corresponding to the first peak of static structure factor and $r_j(t)$ is the position of oxygen atom of water molecule $j$ at a time $t$. The SISFs at several temperatures are shown in **Fig. 3a**. The initial decay for $t < 0.5$ ps, which is approximately fitted by a Gaussian function with respect to time, is due to the ballistic or inertial motion within a local configuration, i.e., a cage. The later part of the decay, called $\alpha$-relaxation, is well fitted with a stretched exponential function and is due to the relaxation of the cage. *The SISFs for the temperature higher than 160 K show complete decays within 80 μs*. It is however that the SISFs decay very slowly below 150 K, with larger standard deviations than those at higher temperatures, but their $\alpha$-relaxation times can still be evaluated reliably.

The relaxation time data obtained by fitting the $\alpha$-relaxation of SISF to the stretched exponential function are fitted with Vogel-Fulcher-Tammann (VFT) equation and plotted in **Fig. 3b**. We can see in the figure that there exist *two* dynamical transitions at ~220 K and ~190 K and three VFT curves are needed to fit whole data for 300 K – 130 K. The dynamical transition at ~220 K has been called a 'fragile-strong' transition due to the change in fragility.

This terminology actually implies that there is a crossover from a marked non-Arrhenius temperature dependence of relaxation time behavior above ~220 K to a weak non-Arrhenius behavior below ~220 K.[9, 18, 22-24, 29, 30, 41, 59, 60] This transition is considered to be caused by the crossover between HDL and LDL by the crossing of the Widom line. As seen in Fig. 2a, the crossover between the locally tetrahedral- and distorted-structured molecules does take place between 220 K and 230 K. The thermodynamic response functions rapidly change in this temperature range (see Fig. 1b). The relaxation time of complex specific heat also yields the transition at ~220 K.[29] The relaxation time data above ~220 K can also be well fitted by a power law $(T-T_C)^{-\gamma}$ (gray curve in Fig. 3c), which is the functional form predicted by the mode coupling theory, as was also observed in the previous studies.[22, 30, 41]

**Figure 3b** shows that the dynamical transition at 190 K is considerably more distinct than that at ~220 K. This second transition takes place where the dynamics change due to the fragmentation of the locally distorted-structured clusters. Thus, a significant change of the thermodynamic response functions as well as the energy and enthalpy are not found at this second transition (see Fig. 1b), unlike the dynamical transition at ~220 K which is associated with the divergence-like behavior of thermodynamic response functions by crossing the Widom line. It is known that the percolation transitions are often accompanied by weak or no thermodynamic anomaly.

It is interesting to note here that Mazza et al. have found a similar, two-dynamical transition scenario, in the dielectric relaxation of a hydrated lysozyme solution. Their careful analysis yielded two transitions, at temperatures ~250 K and ~180 K.[19] By using a coarse-grained model of an adsorbed water monolayer, they assigned the two dynamical transitions to two structural changes of HB network; the first transition at ~250 K is related to the fluctuation of HB formation and breaking (as the change at ~220 K in bulk water) and the second one at ~180 K to that of the ordering of local arrangement of HBs. The model calculation shows that both the transitions are associated with the maxima of specific heat.[19] The specific heat maximum is much distinctive at the second transition at lower temperature. This is different from the present result for bulk water where no such strong thermodynamic anomaly is observed at 190 K. We assume that the relaxation processes of water may be strongly modified with the existence of lysozyme. However, this difference merits further in-depth studies, especially with the hydrated protein where experimentation is possible.

The third curve (below 190 K) of Fig. 3b shows the most Arrhenius-like behavior of the three branches.[34, 61, 62] Therefore, the second dynamical transition at 190 K is to be assigned as a 'strong-extremely strong' transition. As shown below, the supercooled water below 190 K can indeed be characterized as a strong liquid in terms of the Stokes-Einstein relation (see Fig. 5) as well. This characteristic feature of the deeply supercooled water found here is consistent with the experimental result by Amann-Winkel et al.[34] The extrapolation of the third branch below 145 K shows that the relaxation time becomes larger than 100 s at ~135 K roughly assigned as the glass transition temperature. Recent result with the TIP4P/2005 model has also predicted that water is in the glass state below ~140 K.[63]

Our conclusions are based on extensive simulations, namely, five or six sets of the independent equilibration runs and of the subsequent 80 $\mu$s-length production runs at each temperature lower than 190 K and careful analysis of the accumulated data. This led us to observe, for example, the structural relaxation occurring at as low a temperature as 150 K; see Fig. 3a. It is however conceivable that despite all these the system might not be fully equilibrated below ~170 K. Some of the defects, which remain in the present simulations, could be eliminated if much longer equilibration runs were performed. Under such a scenario, the third branch in Fig. 3b would possibly give lower bound estimates for the relaxation times below 170 K. This could be more likely for the lower temperatures, and the temperature dependence of the relaxation time would then show a stronger Arrhenius-like behavior at very low temperatures than captured in the present results. As for below 160 K, much longer simulations must be performed in the future work. However, we expect that the general features of the structural and dynamical transitions observed at ~190 K would not change substantially even in the longer simulations, and the main conclusions of the present study should remain valid.

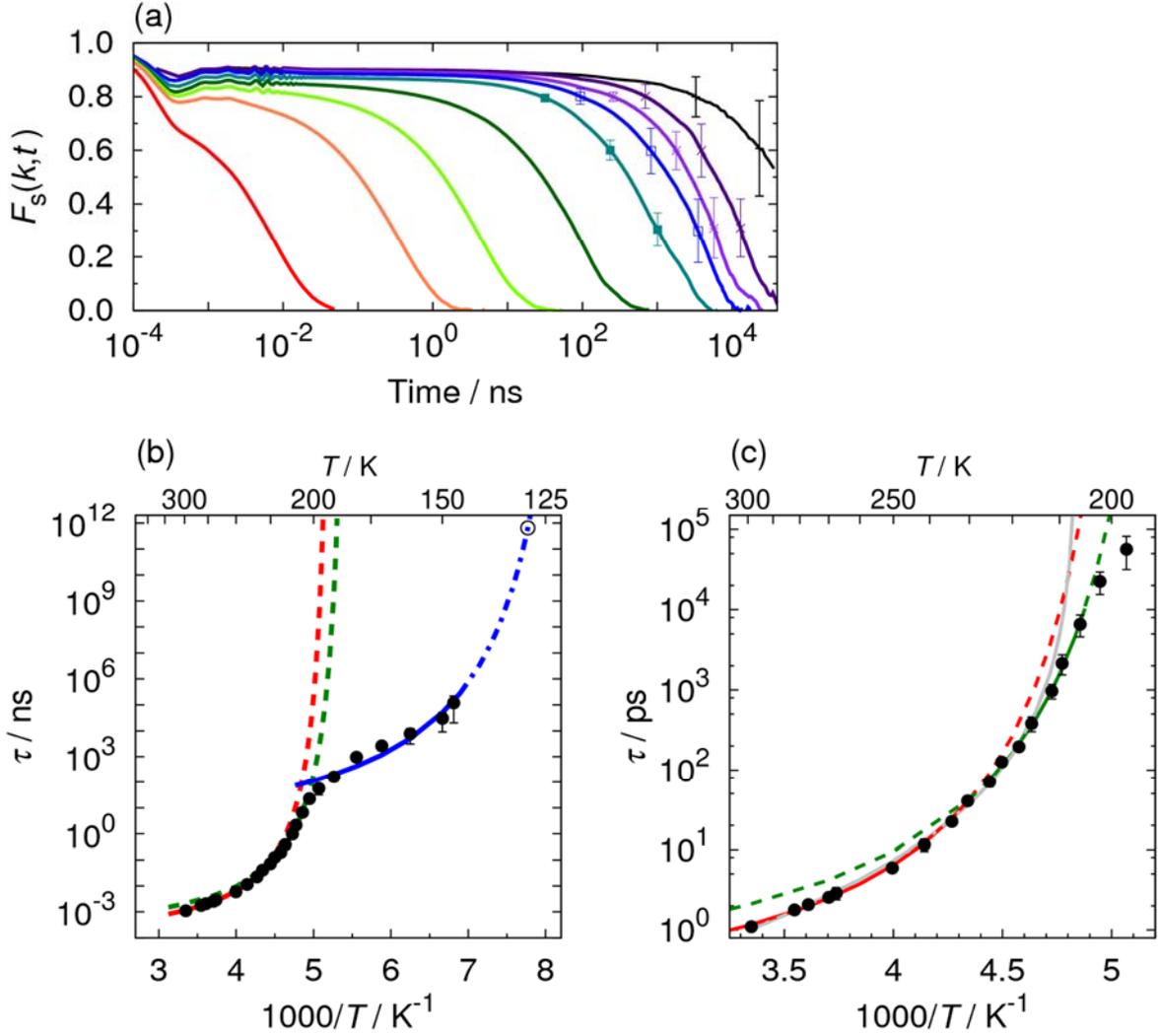

FIG. 3. (a) Self-term of intermediate scattering functions (SISFs) at 250 (red), 220 (orange), 205 (light green), 197 (dark green), 180 (cyan), 170 (blue), 160 (violet), 150 (indigo), and 145 K (black). The standard deviations of SISFs are also shown below 180K. The relaxation is slow with decreasing temperature. (b) Relaxation times from 300 K to 130 K and (c) extended figure at the temperature range between 310 K and 190 K. In (b), the relaxation times are fitted by the Vogel-Fulcher-Tammann (VFT) equation, $\exp(DT_0/(T-T_0))$, and the first, second, and third VFT equations are shown in red, green, and blue lines. The VFT temperature, $T_0$, and the fragility index, $D$, are 188, 181, and 114 K, and 1.25, 1.36, and 3.06, for three regions. Here, the larger fragility index corresponds to the more Arrhenius-like behavior. The relaxation time at 130 K, shown by a white circle, is estimated by extrapolating the relaxation times for 190 K – 145 K. In (c), red and green curves are the fit with the VFT equation for the first and second branches. Gray curve is the fit with a power law, $(T-T_C)^{-\gamma}$. The critical temperature $T_C$ and exponent are 206.3 K and $\gamma=2.63$, respectively. It is noted that all the relaxation times above 190 K cannot be fitted with a single VFT equation though the curvature change at ~220 K from the first ($T > 220$ K) to the second curve (190 K $< T <$ 220 K) is rather weak.

### D. Hydrogen bond defects involved in structural relaxations and low $T_g$

Since the structure and dynamics of water are affected by HB defect molecules,[25, 64] we examine the nature of these defects and their effects on the dynamics and thermodynamics of water. **Figure 4a** shows the temperature dependence of the fractions of three-, four-, and five-coordinated molecules. Above 230 K, liquid water contains many HB defect molecules. The fraction of the HB defect molecules sharply changes at ~220 K, as also observed in the previous study,[25] and become very small below 190 K. Although the defects are very few, they play an important role in the dynamic and thermodynamic properties of water (see below). Moreover, the description of relaxation in terms of defects become increasingly robust as the temperature decreases, in particular below 190 K, because the defects become long lived. The origin of the defects can be both thermodynamic and kinetic, and they are hard to annihilate due to entropic reasons.

The main defects are three- and five-coordinated water molecules. Among the three-coordinated defects there are two types: one is the $H^1O^2$-defect with one HB donor and two HB acceptors and the other is the $H^2O^1$-defect with two HB donor and one-HB acceptors (**Fig. 4b**). The five-coordinated defects mainly consist of the $H^2O^3$-defect with two HB donors and three HB acceptors (Fig. 4b). The present result shows that the $H^1O^2$-defects have a shorter lifetime than the $H^2O^1$- and $H^2O^3$-defects (**Figs. S3d** and **S3e**). Furthermore, we find that the $H^1O^2$- defects have large density of states at 300-400 cm$^{-1}$, which couple with delocalized motions leading to the structural changes (supplementary material and **Fig. S3c**).[65] The analysis of the self-term of van Hove function (supplementary material and **Fig. S3f-S3h**) reveals that these defects indeed induce the dislocations and changes of local water structure, and they themselves sometimes move a distance even larger than 3 Å. In addition to these results, **Fig. 4c** clearly demonstrates that *the decrease in the number of $H^1O^2$-defects correlates well with the growth of the relaxation time.* These results further show that, when compared with the $H^2O^1$- and $H^2O^3$-defects, the $H^1O^2$-defects play a more central role as a promoter of structural changes, inducing facile rotational and translational motions in water (**Fig. S4**). Note that the $H^1O^2$-defects have the characteristics of a Glarum defect well-known in the dielectric relaxation of glasses[66], since the initial defect becomes a four-coordinated molecule and the neighboring four-coordinated molecule is transformed into a new three-coordinated defect.

**Figure 4d** depicts the temperature dependence of the fraction of the HB defects in the inherent structures of water. The population of the defects rapidly decreases with lowering temperature below ~220 K, but changes slowly below 190 K. The $H^2O^1$- and $H^2O^3$-defects surrounded by the firm HB network formed by locally tetrahedral-structured molecules survive even below 190 K and are hardly annihilated, whereas the number of $H^1O^2$-defects diminishes to zero at ~135 K. As shown in Fig. 4c, the number of $H^1O^2$-defects strongly correlates with the relaxation time, so that the presence of this defects is a key element for sustaining the molecular motions in the third blanch. *In other words, these defects serve as the harbinger of the impending glass transition of water. Moreover, its presence serves to lower $T_g$ of water to $T_g/T_m \approx 1/2$, instead of the empirical rule of $T_g/T_m \approx 2/3$*[35] ($T_m \approx 252$K for the TIP4P/2005 model[39]). Note that if the fraction of the $H^1O^2$-defects above 220 K is extrapolated to low temperatures by assuming that the third branch never intervenes (dashed-dotted violet curve in Fig. 4d), the number of the $H^1O^2$-defects becomes zero at ~180 K, *i.e., water would have a glass transition temperature as expected from the 2/3 rule.*[35]

It has been proposed recently that quantum effects, such as tunneling effect, decrease the value of $T_g$[67] and the ratio $T_g/T_m$ of water.[68] The quantum effects on the structural and dynamical properties of water are non-negligible[69] and the isotope effects on the properties of water have been discussed.[70, 71] However, experimentally $T_g/T_m$ of $D_2O$ is 0.53, very close to $T_g/T_m$ of $H_2O$,[68] which shows that there is only a small isotope effect on $T_g/T_m$. Therefore, the ratio of $T_g/T_m \approx 1/2$ is mainly attributed to the presence of the third branch below 190 K.

*It is interesting to note that this active three-coordinated $H^1O^2$-defect is the same kind of defect which promotes melting of ice.*[72] Since the $H^1O^2$-defects easily induce the structural changes by exchanging HBs with neighboring four-coordinated water molecules, they can be the primal moving defect in the relaxation of the highly viscous LDL between 136 K and 150 K. Recently, a second glass transition has been found between HDL and HDA at a very low temperature, ~116 K.[34] It would be of great interest to examine the role of $H^1O^2$-defects in the second glass transition, because the low glass transition temperature of HDL with a larger concentration of the defects than LDL is consistent with the picture developed here.

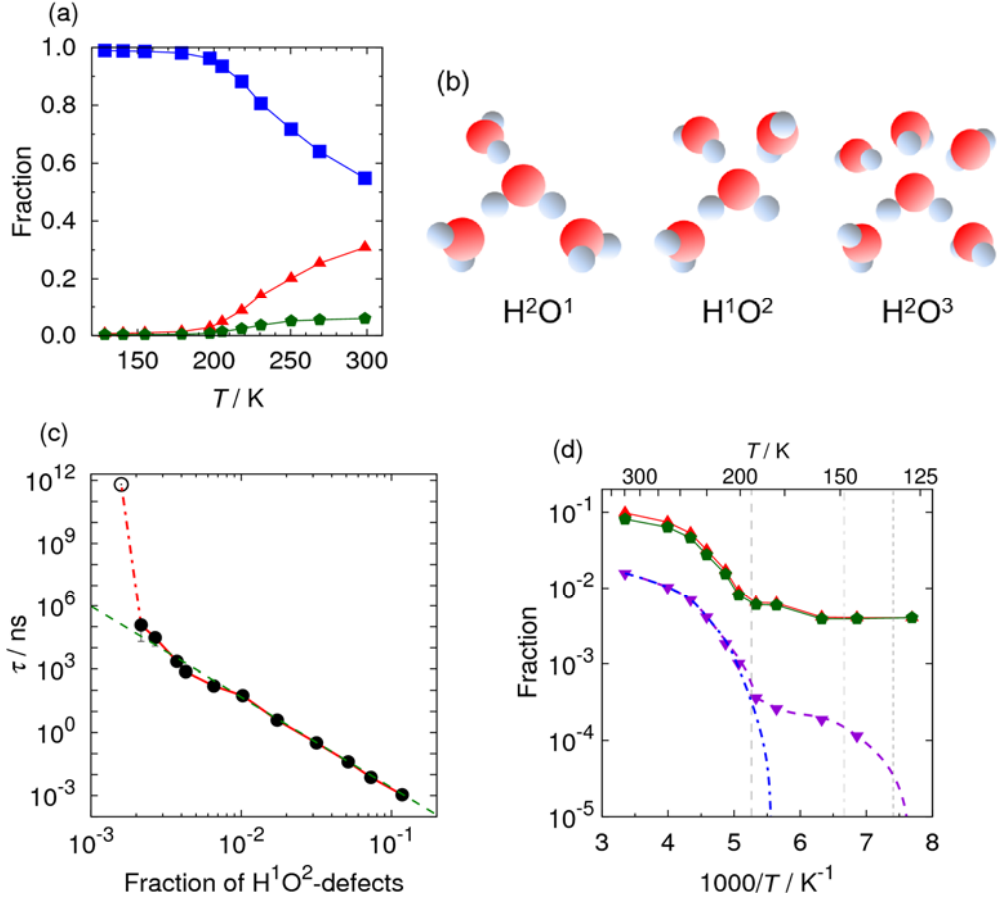

FIG. 4. (a) Fractions of three- (red), four- (blue), and five- (green) coordinated water molecules, (b) schematic figures of $H^2O^1$- (left), $H^1O^2$- (center), and $H^2O^3$- (right) defects, (c) the dependence of relaxation time of self-term of intermediate scattering function (Eq. (1)) on the fraction of $H^1O^2$-defects, and (d) the fractions of $H^2O^1$- (red), $H^1O^2$- (violet), and $H^3O^2$- (green) defects. In (b), the $H^1O^2$-defect easily induces HB structural changes because the fourth nearest neighbor molecule (not shown) is not located at an expected tetrahedral position. In (c), it is noted that the relaxation time is approximately proportional to $(N_{H^1O^2})^{-\alpha}$ ($\alpha=4.3$) (dashed green line) for temperatures down to ~145K. In (d), the defects are defined in inherent structures, which are potential energy minima obtained from the instantaneous structures. Dashed, dotted-dashed, and dotted vertical lines mark 190, 150, and 135 K, respectively. Note that, at very low temperatures, the fraction of $H^1O^2$-defects goes to zero and the numbers of remaining three- and five-coordinated defects become identical, as they form stable defect pairs. Note that the fraction of $H^1O^2$-defects decays in two distinct stages: The first curve describes the temperature dependence down to 190 K and the second one below 190 K. The second curve decays to zero at ~130 K, whereas the extension of fraction of $H^1O^2$-defects down to 220 K decays to zero at ~180 K as seen in blue dashed-dotted curve. The temperature dependence of fraction of locally distorted-structured molecules also diminishes at ~180 K if its curve in the same temperature range is extrapolated to lower temperatures. All these results show that water should undergo the glass transition at ~180 K if the fragmentation of locally distorted-structured clusters, i.e., the third branch, does not intervene.

**E. Change in fragility and the Stokes-Einstein relation**

The remarkable change in fragility at 190 K suggests a change in the coupling of transport coefficients to the collective dynamics, like density fluctuations. This can be probed by the Stokes-Einstein (SE) relation. The SE relation is the relationship between the diffusion constant, $D$, and the temperature-scaled relaxation time of the self-term of intermediate scattering function, $\tau_{ISF}^{self}/T$. Such linearity in the SE relation, $D \sim (\tau_{ISF}^{self}/T)^{-1}$, has indeed been observed in the experimental result of the confined supercooled water in the MCM-41-S nanotubes above 290 K[73] and in the simulation of water with ST2 model above 275 K[74] and the TIP5P model above 320 K.[73] On the other hand, a fractional exponent ($\zeta$) dependence of the SE relation, i.e., $D \sim (\tau_{ISF}^{self}/T)^{-\zeta}$ ($\zeta < 1$), has been found in the experimental result of the confined water ($\zeta = 0.62$ from 290 K to ~180 K),[73] and in the simulation results ($\zeta = 0.84$ from 350 K to 210 K with the SPC/E model,[75] $\zeta = 0.77$ from 275 K to 250 K with the ST2 model,[74] and $\zeta = 0.77$ from 320 K to ~200 K with the TIP5P model[73]). As seen in **Fig. 5**, the present result shows a similar fractional $\zeta$ dependence of the SE relation, $D \sim (\tau_{ISF}^{self}/T)^{-\zeta}$ ($\zeta \approx 0.74$), above 190 K. The breakdown of SE relation may be attributed to the dynamic heterogeneity arising from the interconversion of locally tetrahedral- and distorted-structured molecules.

The present result shows, however, that the SE relation is almost recovered below 190 K when the diffusion coefficient behaves as $D \sim (\tau_{ISF}^{self}/T)^{-\zeta}$ with $\zeta \approx 0.95$. This result can be explained as follows: below 190 K, the fraction of the locally distorted-structured molecules is small and the clusters of locally distorted-structured molecules are fragmented into small clusters in the firm network of locally tetrahedral-structured molecules (Figs. 2b and 2f), and thus while the interconversion dynamics between locally tetrahedral- and distorted-structured molecules exist but they are significantly suppressed below 190 K. The present results above and below 190 K are consistent with theoretical results for both fragile liquids with $\zeta \approx 0.73$ and strong liquids with $\zeta \approx 0.95$.[76] Furthermore, the present result of $\zeta$-dependence below 190 K is consistent with the theoretical model for strong liquids,[77] and the analysis of fragility (Fig. 3b). The present outcome of a strong liquid at low temperatures is consistent with the experimental result on the deeply supercooled water between 136 K and 150 K.[34]

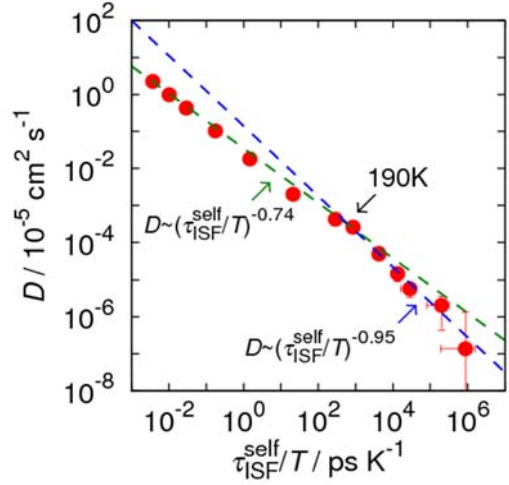

FIG. 5. Stokes-Einstein relation. $D$ and $\tau_{ISF}^{self}$ are the diffusion constant and the relaxation time of self-term of intermediate scattering function, respectively. The slopes of green and blue lines are -0.74 and -0.95, respectively.

## IV. CONCLUSIONS

The present study revealed that the vitrification of water occurs at least in three distinct stages accompanied by two changes in fragility. In contrast to the first dynamical transition at 220 K associated with the crossing of the Widom line, the newly found second transition at 190 K was found to occur by the fragmentation and isolation of the locally distorted-structured clusters. Detailed, long simulations undertaken in this study showed that below 190 K the locally distorted-structured molecules exist as the dispersed three- and five-coordinated defects in the locally tetrahedral-structured network. We further clarified the role of the $H^1O^2$-defects; these defects induce structural relaxations even in the firm HB network of the supercooled water. The present study demonstrated, for the first time, the existence of these defects, scattered as they are in the continuous HB network and hard to annihilate, are responsible for the significant reduction of the glass transition temperature of water. It is the fragmentation and isolation of locally distorted-structured clusters at 190 K and the hitherto unknown role of the active $H^1O^2$-defect that serve to explain many of the anomalies in deeply supercooled region of liquid water, such as the limited existence of the observed supercooled liquid state between 136 K and 150 K. This role of defects unraveled here will provide a deep understanding of the myriad of anomalous properties of supercooled water.

Further studies are needed (1) to explore the properties of water at high and negative pressures; for example, the pressure dependences of the cluster fragmentations and the three relaxation processes, and (2) to develop a more sophisticated model potential of water molecules including such as many-body effects and perform much longer trajectory calculations with it. The fragmentation and isolation mechanism of the locally distorted-structured clusters and the role of the active $H^1O^2$-defect should be confirmed in such calculations.

## SUPPLEMENTARY MATERIAL

See supplementary material for the table of simulation details and the analyses and figures of structural relaxations, local liquid structure, hydrogen bond defects, non-Gaussian parameter, and course-grained dynamics.

## ACKNOWLEDGMENTS

The present study was supported by JSPS, Grant-in Aid for Scientific Research (A) (JH16H02254) for S.S., by JC Bose Fellowship (DST, India) for B. B., and Indo (DST)-Japan

# REFERENCES

1. D. Eisenberg and W. Kauzmann, *The Structure and Properties of Water*. (Oxford Univ. Press, Oxford, 1969).
2. G. W. Z. Robinson, S. B.; Singh, S.; Evans, M. W., *Water in Biology, Chemistry, and Physics: Experimental Overviews and Computational Methodologies*. (World Scientific, Singapore, 1996).
3. O. Mishima and H. E. Stanley, Nature **396**, 329-335 (1998).
4. P. G. Debenedetti, J. Phys. Condens. Matter. **15**, R1669-R1726 (2003).
5. R. J. Speedy, J. Phys. Chem. **86**, 982-991 (1982).
6. P. H. Poole, F. Sciortino, U. Essman and H. E. Stanley, Nature **360**, 329-335 (1992).
7. S. Sastry, P. G. Debenedetti, F. Sciortino and H. E. Stanley, Phys. Rev. E **53**, 6144-6154 (1996).
8. D. Paschek and A. Geiger, J. Chem. Phys. B **103**, 4139-4146 (1999).
9. C. A. Angell, Science **319**, 582-587 (2008).
10. E. B. Moore and V. Molinero, Nature **479**, 506-508 (2011).
11. J. C. Palmer, F. Martelli, Y. Liu, R. Car, A. Z. Panagiotopoulos and P. G. Debenedetti, Nature **510**, 385-388 (2014).
12. A. Nilsson and L. G. Pettersson, Nat. Commun. **6**, 8998 (2015).
13. P. Gallo, K. Amann-Winkel, C. A. Angell, M. A. Anisimov, F. Caupin, C. Chakravarty, E. Lascaris, T. Loerting, A. Z. Panagiotopoulos, J. Russo, J. A. Sellberg, H. E. Stanley, H. Tanaka, C. Vega, L. Xu and L. G. Pettersson, Chem. Rev. **116**, 7463-7500 (2016).
14. P. H. Handle, T. Loerting and F. Sciortino, Proc. Natl. Acad. Sci. U S A **114**, 13336-13344 (2017).
15. J. C. Palmer, A. Haji-Akbari, R. S. Singh, F. Martelli, R. Car, A. Z. Panagiotopoulos and P. G. Debenedetti, J. Chem. Phys. **148**, 137101 (2018).
16. A. Manka, H. Pathak, S. Tanimura, J. Wolk, R. Strey and B. E. Wyslouzil, Phys. Chem. Chem. Phys. **14**, 4505-4516 (2012).
17. J. A. Sellberg, C. Huang, T. A. McQueen, N. D. Loh, H. Laksmono, D. Schlesinger, R. G. Sierra, D. Nordlund, C. Y. Hampton, D. Starodub, D. P. DePonte, M. Beye, C. Chen, A. V. Martin, A. Barty, K. T. Wikfeldt, T. M. Weiss, C. Caronna, J. Feldkamp, L. B. Skinner, M. M. Seibert, M. Messerschmidt, G. J. Williams, S. Boutet, L. G. Pettersson, M. J. Bogan and A. Nilsson, Nature **510**, 381-384 (2014).
18. C. E. Bertrand, Y. Zhang and S. H. Chen, Phys. Chem. Chem. Phys. **15**, 721-745 (2013).
19. M. Mazza, K. Stokely, S. E. Pagnotta, F. Bruni, H. E. Stanley and G. Franzese, Proc. Natl. Acad. Sci. USA **108**, 19873-19878 (2011).
20. M. Yamada, S. Mossa, H. E. Stanley and F. Sciortino, Phys. Rev. Lett. **88**, 195701 (2002).
21. D. Paschek, Phys. Rev. Lett. **94**, 217802 (2005).
22. L. Xu, P. Kumar, S. V. Buldyrev, S. H. Chen, P. H. Poole, F. Sciortino and H. E. Stanley, Proc. Natl. Acad. Sci. U S A **102**, 16558-16562 (2005).

23. D. Corradini, M. Rovere and P. Gallo, J. Chem. Phys. **132**, 134508 (2010).
24. J. L. Abascal and C. Vega, J. Chem. Phys. **133**, 234502 (2010).
25. P. H. Poole, S. R. Becker, F. Sciortino and F. W. Starr, J. Phys. Chem. B **115**, 14176-14183 (2011).
26. T. Yagasaki, M. Matsumoto and H. Tanaka, Phys. Rev. E **89**, 020301 (2014).
27. T. Sumi and H. Sekino, RSC Advances **3**, 12743-12750 (2013).
28. K. T. Wikfeldt, A. Nilsson and L. G. Pettersson, Phys. Chem. Chem. Phys. **13**, 19918-19924 (2011).
29. S. Saito, I. Ohmine and B. Bagchi, J. Chem. Phys. **138**, 094503 (2013).
30. M. De Marzio, G. Camisasca, M. Rovere and P. Gallo, J. Chem. Phys. **144**, 074503 (2016).
31. O. Mishima, L. D. Calvert and E. Whalley, Nature **310**, 393-395 (1984).
32. O. Mishima, L. D. Calvert and E. Whalley, Nature **314**, 76-78 (1985).
33. T. Loerting, C. Salzmann, I. Kohl, E. Mayer and A. Hallbrucker, Phys. Chem. Chem. Phys. **3**, 5355-5357 (2001).
34. K. Amann-Winkel, C. Gainaru, P. H. Handle, M. Seidl, H. Nelson, R. Bohmer and T. Loerting, Proc. Natl. Acad. Sci. USA **110**, 17720-17725 (2013).
35. W. Kauzmann, Chem. Rev. **43**, 219-256 (1948).
36. G. P. Johari, A. Hallbrucker and E. Mayer, Nature **330**, 552-553 (1987).
37. R. S. Smith and B. D. Kay, Nature **398**, 788-791 (1999).
38. M. J. Abraham, T. Murtola, R. Schulz, S. Páll, J. C. Smith, B. Hess and E. Lindahl, SoftwareX **1-2**, 19-25 (2015).
39. J. L. F. Abascal and C. Vega, J. Chem. Phys. **123**, 234505 (2005).
40. H. L. Pi, J. L. Aragones, C. Vega, E. G. Noya, J. L. F. Abascal, M. A. Gonzalez and C. McBride, Mol. Phys. **107**, 365-374 (2009).
41. F. W. Starr, F. Sciortino and H. E. Stanley, Phys. Rev. E **60**, 6757-6768 (1999).
42. G. Franzese and H. E. Stanley, J. Phys. Condens. Matter **19**, 205126 (2007).
43. P. Beaucage and N. Mousseau, J. Phys. Condens. Matter **17**, 2269-2279 (2005).
44. A. Haji-Akbari and P. G. Debenedetti, Proc. Nat. Acad. Sci. USA **112**, 10582-10588 (2015).
45. A. Haji-Akbari and P. G. Debenedetti, Proc. Natl. Acad. Sci. USA **114**, 3316-3321 (2017).
46. J. L. Abascal, E. Sanz, R. Garcia Fernandez and C. Vega, J. Chem. Phys. **122**, 234511 (2005).
47. J. R. Espinosa, C. Navarro, E. Sanz, C. Valeriani and C. Vega, J. Chem. Phys. **145**, 211922 (2016).
48. F. Mallamace, C. Branca, M. Broccio, C. Corsaro, C. Y. Mou and S. H. Chen, Proc. Natl. Acad. Sci. USA **104**, 18387-18391 (2007).
49. V. Holten, J. C. Palmer, P. H. Poole, P. G. Debenedetti and M. A. Anisimov, J. Chem. Phys. **140**, 104502 (2014).
50. J. Russo and H. Tanaka, Nat. Commun. **5**, 3556-3566 (2014).

51. R. S. Singh, J. W. Biddle, P. G. Debenedetti and M. A. Anisimov, J. Chem. Phys. **144** (14), 144504 (2016).
52. J. W. Biddle, R. S. Singh, E. M. Sparano, F. Ricci, M. A. Gonzalez, C. Valeriani, J. L. Abascal, P. G. Debenedetti, M. A. Anisimov and F. Caupin, J. Chem. Phys. **146**, 034502 (2017).
53. A. Luzar and D. Chandler, J. Chem. Phys. **98**, 8160-8173 (1993).
54. P. Raiteri, A. Laio and M. Parrinello, Phys. Rev. Lett. **93**, 087801 (2004).
55. E. B. Moore and V. Molinero, J. Chem. Phys. **130**, 244505 (2009).
56. H. E. Stanley and J. Teixeira, J. Chem. Phys. **73**, 3404-3422 (1980).
57. S. Banerjee, R. Ghosh and B. Bagchi, J Phys Chem B **116**, 3713-3722 (2012).
58. P. Bhimalapuram, S. Chakrabarty and B. Bagchi, Phys. Rev. Lett. **98**, 206104 (2007).
59. A. Faraone, L. Liu, C. Y. Mou, C. W. Yen and S. H. Chen, J. Chem. Phys. **121**, 10843-10846 (2004).
60. L. Liu, S. H. Chen, A. Faraone, C. W. Yen and C. Y. Mou, Phys. Rev. Lett. **95**, 117802 (2005).
61. K. Ito, C. T. Moynihan and C. A. Angell, Nature **398**, 492-495 (1999).
62. C. A. Angell, Chem. Rev. **102**, 2627-2650 (2002).
63. J. Wong, D. A. Jahn and N. Giovambattista, J. Chem. Phys. **143**, 074501 (2015).
64. F. Sciortino, A. Geiger and H. E. Stanley, Nature **354**, 218-221 (1991).
65. I. Ohmine and H. Tanaka, J. Chem. Phys. **93**, 8138-8147 (1990).
66. S. H. Glarum, J. Chem. Phys. **33**, 639-643 (1960).
67. V. N. Novikov and A. P. Sokolov, Phys. Rev. Lett. **110**, 065701 (2013).
68. C. Gainaru, A. L. Agapov, V. Fuentes-Landete, K. Amann-Winkel, H. Nelson, K. W. Köster, A. I. Kolesnikov, V. N. Novikov, R. Richert, R. Böhmer, T. Loerting and A. P. Sokolov, Proc. Natl. Acad. Sci. USA **111**, 17402-17407 (2014).
69. R. A. Kuharski and P. J. Rossky, Chem. Phys. Lett. **103**, 357-362 (1984).
70. L. Hernández de la Peña and P. G. Kusalik, J. Amer. Chem. Soc. **127**, 5246-5251 (2004).
71. L. Wang, M. Ceriotti and T. E. Markland, J. Chem. Phys. **141**, 104502 (2014).
72. K. Mochizuki, M. Matsumoto and I. Ohmine, Nature **498**, 350-354 (2013).
73. L. Xu, F. Mallamace, Z. Yan, F. W. Starr, S. V. Buldyrev and H. E. Stanley, Nat. Phys. **5**, 565-569 (2009).
74. S. R. Becker, P. H. Poole and F. W. Starr, Phys. Rev. Lett. **97**, 055901 (2006).
75. M. G. Mazz, N. Giovambattista, H. E. Stanley and F. W. Starr, Phys Rev E **76**, 031203 (2007).
76. Y. Jung, J. P. Garrahan and D. Chandler, Phys. Rev. E **69**, 061205 (2004).
77. A. J. Moreno, S. V. Buldyrev, E. La Nave, I. Saika-Voivod, F. Sciortino, P. Tartaglia and E. Zaccarelli, Phys. Rev. Lett. **95**, 157802 (2005).